\documentclass[fleqn,10pt]{wlscirep}
\title{Nonlinear network model analysis of vibrational energy transfer and localisation in the Fenna-Matthews-Olson complex}

\author[1,*]{Sarah E. Morgan}
\author[1,2]{Daniel J. Cole}
\author[1]{Alex W. Chin}
\affil[1]{Theory of Condensed Matter Group, Physics Department, University of Cambridge, CB3 0HE, United Kingdom}
\affil[2]{School of Chemistry, Newcastle University, Newcastle upon Tyne, NE1 7RU, United Kingdom}

\affil[*]{sem91@cam.ac.uk}
\usepackage[colorinlistoftodos]{todonotes}
\usepackage{mathpazo}

\begin{abstract}
Collective protein modes are expected to be important for facilitating energy transfer in the Fenna-Matthews-Olson (FMO) complex, however to date little work has focussed on the microscopic details of these vibrations. The nonlinear network model (NNM) provides a computationally inexpensive approach to studying vibrational modes at the microscopic level, whilst incorporating anharmonicity in the inter-residue interactions which can influence protein dynamics. We apply the NNM to the FMO complex and find evidence for the existence of nonlinear discrete breather modes. These modes tend to transfer energy to the highly connected core pigments, potentially opening up alternative excitation energy transfer routes. Incorporating localised modes based on these discrete breathers in the optical spectra calculations for FMO using ab initio site energies and excitonic couplings can substantially improve their agreement with experimental results.

\end{abstract}
\begin{document}

\flushbottom
\maketitle

\thispagestyle{empty}

%\noindent Please note: Abbreviations should be introduced at the first mention in the main text – no abbreviations lists. Suggested structure of main text (not enforced) is provided below.

\section*{Introduction}

%The Introduction section, of referenced text\cite{Figueredo:2009dg} expands on the background of the work (some overlap with the Abstract is acceptable). The introduction should not include subheadings.

Recent 2D electronic spectroscopy experiments have provided evidence for long-lasting high frequency oscillations in light-harvesting pigment-protein complexes \cite{Engel2007,Collini2010,Fleming2011,Ogilvie2012}. These oscillations can arise from either ground state vibrations or electronic coherences, the latter being controversial due to the long dephasing times observed which can be on the picosecond timescale \cite{Panitchayangkoon2010}. There is growing evidence that vibrational modes play an important role and may mix with electronic states to produce mixed vibrational/electronic signals \cite{Fassioli2010, Christensson2012,Jonas2013,Chin2012,Butkus2013,Thorwart2015}. However even intramolecular vibrations would be expected to decay considerably on a picosecond timescale and debates about the origins of these oscillations are ongoing. In addition to high frequency intramolecular pigment modes, lower frequency collective protein modes are also expected to be important \cite{Fassioli2010,Jonas2013,Chin2012}. Protein vibrations can remain out of equilibrium over long time periods \cite{Hu2016} and tend to be much less rigid than intramolecular pigment modes. Hence non-linear effects are often important \cite{Levy1982,Hayward1995} and could provide alternative, intriguing energy transfer mechanisms. Long-lasting modes have also been observed in IR experiments \cite{Xie2002,Acbas2014}, for example oscillations which last for over 500 ps in bacteriorhodopsin \cite{Xie2002}. Further work is needed to build a microscopic description of these vibrations and to clarify their role in biological light harvesting complexes.

The nonlinear network model proposed by Juanico et al. \cite{Juanico2007} offers a promising approach to studying nonlinear protein dynamics in a computationally inexpensive way. Interestingly, previous studies applying this model to a number of proteins have found the spontaneous localization of energy and formation of discrete breather modes (DBs) \cite{Piazza2008,Luccioli2011,Caraglio2014}. These localised vibrational modes are able to harvest energy from their surroundings and to transfer energy between different protein residues on picosecond timescales. If supported by light harvesting pigment protein complexes, DBs could therefore be functionally important for energy transfer and storage \cite{Tsironis2001}. Whilst most work on DBs to date is theoretical, correlated motions in ubiquitin revealed by experimental NMR measurements match the calculated displacement patterns of DBs \cite{Piazza2014}, suggesting that simple DB models are able to capture some experimental features.

Here we investigate the high frequency protein vibrations in a nonlinear network model of an archetypal light-harvesting system, the Fenna-Matthews-Olson complex (FMO). FMO is found in green sulfur bacteria and acts as a funnel, transferring energy from the light harvesting antennae to the reaction centre. It has a trimeric structure in which each monomer contains 8 bacteriochlorophyll-a pigments (BChla) \cite{Fenna1975}. Here we focus on the 7 pigments within the protein envelope, which are shown in Fig. \ref{highestNM} a. FMO has attracted considerable attention recently due to the long-lasting oscillations which have been observed in 2D spectroscopy of the complex even at room temperature \cite{Savikhin1997,Engel2007,Panitchayangkoon2010}.

We begin by studying the high frequency normal modes (NMs) of FMO, several of which are delocalised across the trimer and have strong components on the pigments in the core of the FMO structure. We note that Renger et al. have previously considered the normal modes of FMO (using an atomistic model) \cite{Renger2012}, however their work focussed on deriving the spectral function and did not discuss the spatial properties of individual modes or anharmonic effects. Moreover, because they used an all-atom model they were only able to study a single monomer. Whilst there is only weak excitonic coupling between the monomers, we will show that the trimeric structure can be important for mechanical effects. We then compute the dynamics which follow excitation of the highest frequency NM and find evidence for the existence of DBs in the structure. The DB which forms following excitation of the highest frequency NM is localised close to pigments 3, 4 and 7 and is therefore most likely to modulate the site energies of those pigments. Finally, we study the effect which this localised DB mode might have on the optical spectra; in particular the linear absorption (LA), linear dichroism (LD) and circular dichroism (CD) spectra. To do this, we use the pigment energies and couplings obtained by Cole et al. \cite{Cole2012} from first principles using large scale quantum-mechanical calculations. We then approximate the DB with a mode of frequency $\omega=100~\mathrm{or}~180~ \mathrm{cm}^{-1}$, which couples to pigments 3, 4 and 7. We show that the mode can substantially improve the agreement between the calculated and experimental LA and LD spectra.

\section*{Methods}

We model FMO using the nonlinear network model (NNM) introduced by Juanico et al. \cite{Juanico2007}. In this model each amino acid is represented by a point-like node of mass 110 atomic mass units, which is placed at the position of the corresponding $\mathrm{C}_{\alpha}$ atom. Calculations were based on the holo (8 BChla per monomer) form of the trimeric 1.3~\AA{} X-ray crystal structure of \textit{Prosthecochloris aestuarii} (PDB: 3EOJ)  \cite{Tronrud2009}, although as mentioned above we only include 7 pigments in our model. Based on their relative masses, the chlorin ring and phytol tail of the bacteriochlorophyll a (BChla) pigment molecules were modelled by 5 nodes and 3 nodes, respectively. The nodes were placed at the positions of the MG, C2A, C2B, C2C, C2D, C2, C10 and C18 atoms (crystal structure atom labelling). Together the residues and pigment atoms make up the 1242 nodes of our model.

The NNM potential energy has both a linear and a nonlinear term and reads \cite{Juanico2007,Piazza2014}:

\begin{equation}
  E = \sum_{i>j} c_{ij} \left[ \frac{k_2}{4} \left(r_{ij}- R_{ij}\right)^2 + \frac{k_4}{8} \left( r_{ij}- R_{ij} \right)^4 \right]
	\label{eqn:PE}
\end{equation}

\noindent
where  $r_{ij}= \left|\vec{r}_i- \vec{r}_j\right|$ is the instantaneous distance between nodes i and j and $R_{ij}=\left| \vec{R}_i-\vec{R}_j \right|$ is the distance between nodes i and j at equilibrium (t=0). $c_{ij}=1$ if $R_{ij}$ is less than a cut-off distance, $R_c$, otherwise $c_{ij}=0$. Following the literature, we set $R_c= 10~\text{\AA}$, $k_2=10~ \mathrm{kcal/mol/}\text{\AA}^2$ and $k_4=10~\mathrm{kcal/mol/}\text{\AA}^4$ \cite{Juanico2007, Piazza2014} unless otherwise stated.

We obtain the normal modes of the system by calculating the eigenvectors and eigenvalues of the Hessian matrix. We then perform molecular dynamics simulations using a Verlet algorithm with a 1 fs time step. The network is excited with total energy $E_0$ in the direction of one of the eigenvectors (as specified in the text). We record the total energy of every residue at each time step, given by a sum of the kinetic and potential energies.

In order to calculate the optical spectra, we use the pigment energies and couplings obtained by Cole et al. using large-scale quantum mechanical calculations \cite{Cole2012}. We also couple pigments 3, 4 and 7 to a single vibrational mode in order to represent the observed DBs, which is incorporated on an equal footing with the electronic energy levels. The total Hamiltonian reads:

\begin{equation}
H= \sum_i \epsilon_i \left|i\right\rangle \left\langle i \right| + \sum_{i \neq j} J_{ij} \left|i\right\rangle \left\langle j \right| + \sum_i g_i \left(a + a^{\dagger} \right) \left|i\right\rangle \left\langle i \right| + \hbar \omega a^{\dagger} a
\label{eq:Hamiltonian}
\end{equation}

\noindent
where $\epsilon_i$ is the site energy of pigment i, $J_{ij}$ is the coupling between optical excitations in pigments i and j, $\omega$ is the mode frequency and $g_i$ is the coupling of pigment i to the vibrational mode. $g_i$ depends on the Huang Rhys factor, S, according to $g_i=\sqrt{S}\omega$. To calculate the spectra we use the Master equation approach taken by Marcus \cite{Renger2002} and Cole et al. \cite{Cole2012}, which makes the Markov approximation and hence each excitonic transition has a Lorentzian lineshape. We use the same parameters as Cole et al. for the pure dephasing rate and the spectral density function (which provides a smooth background bath) \cite{Cole2012}. To include the effect of static disorder, the final spectra are obtained by averaging over 1000 realisations of the site energies, each drawn from a Gaussian distribution with a full-width half maximum of $100~\mathrm{cm}^{-1}$. We include 15 harmonic oscillator levels for the mode, which gives converged spectra.

\section*{Results and Discussion}

\begin{figure}[ht]
\centering
\includegraphics[width=150mm]{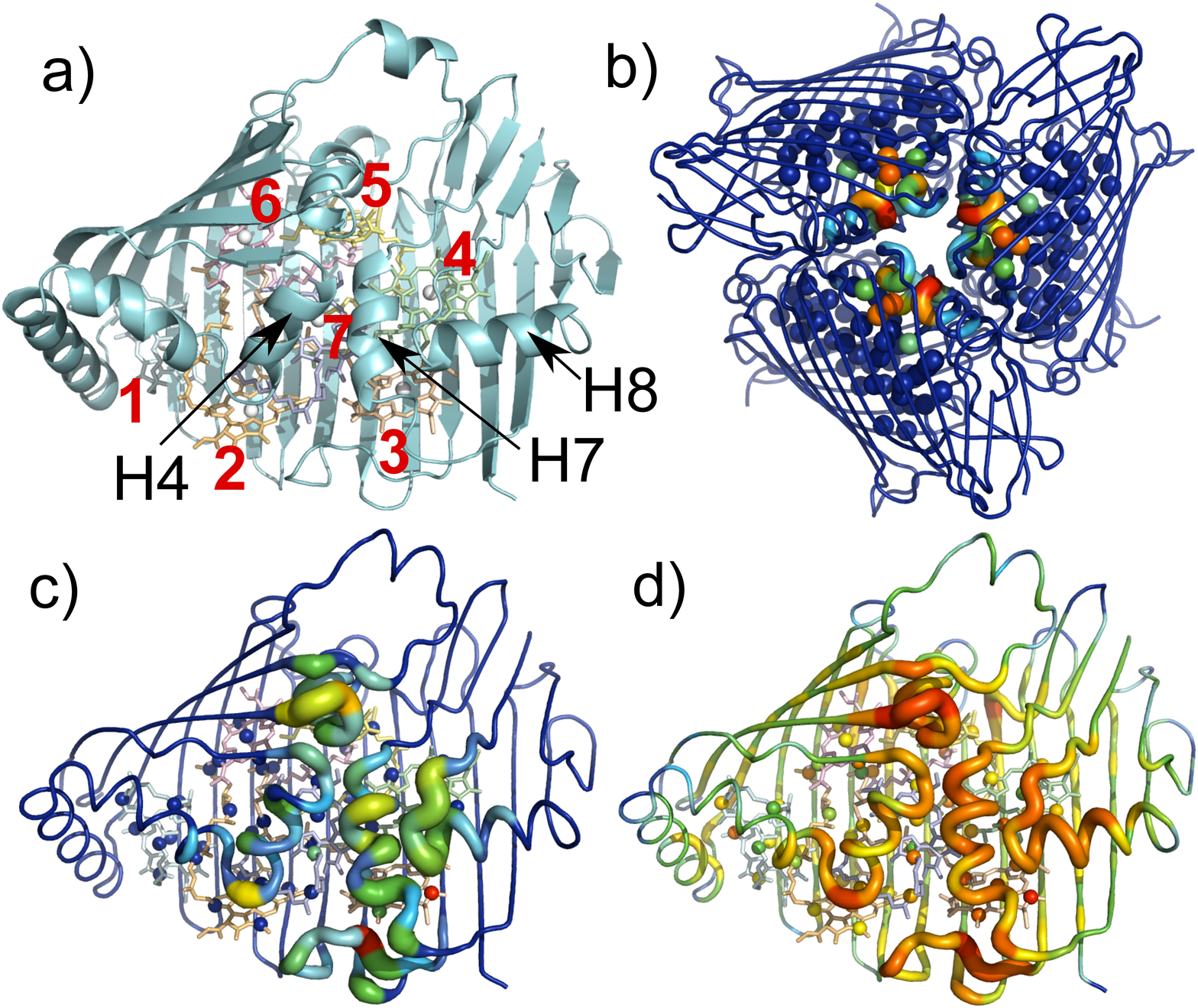}
\caption{\textbf{a)} FMO monomer with the seven pigments and helices 4, 7 and 8 labelled (H4, H7 and H8 respectively). \textbf{b)} FMO trimer coloured according to participation in the highest frequency NM. \textbf{c)} FMO monomer coloured according to average participation in the 20 highest frequency NMs. \textbf{d)} FMO monomer coloured according to node degree. Note that for c) monomer 1 is shown although monomers 2 and 3 are the same to within 0.025\% of the total amplitude whilst for d) all monomers are identical.}
\label{highestNM}
\end{figure}

We begin by using the nonlinear network model of FMO outlined in the methods section to compute the high frequency NMs. Several of these NMs are delocalised across the trimer, for example the highest frequency NM is split equally across the three monomers, with almost identical contributions on individual residues in each monomer (to within $0.03$\% of the total amplitude). The amplitudes squared of the highest frequency NM on each node are plotted in Fig. \ref{highestNM} b and the largest component on any single node is localised on residue LYS 354 (3.6\% of the total amplitude is found on LYS 354 in each monomer). LYS 354 lies on helix 8 and forms an inter-monomer salt bridge with ASP 306 and close hydrophobic contact with PHE 304, which is close to helix 7 on the neighbouring monomer. There are also large components on the pigments, the largest being the 7.4\% localised on pigment 3 (split across the 8 pigment nodes). The amplitude of the NM squared on each node averaged over the 20 highest frequency NMs is shown in Fig. \ref{highestNM} c. The node degree (i.e. the total number of connections each node makes) is plotted in Fig. \ref{highestNM} d. We note that the total number of connections made by pigment atoms for pigments found in the core of the complex (pigments 3, 4, 5, 6 and 7) are higher than those found near the surface of the protein (pigments 1 and 2). Interestingly, in general the high frequency NMs tend to have strong  amplitudes on nodes with high connectivities; Supplementary Fig. 1 plots the amplitude of the NM squared on a node averaged over the 20 highest frequency NMs as a function of the node's degree. Since DBs emerge from the high frequency NMs, this is in keeping with the result of Piazza et al. that highly connected protein regions are always the areas in which DBs form \cite{Piazza2008}.

\begin{figure}[ht]
\centering
\includegraphics[width=85mm]{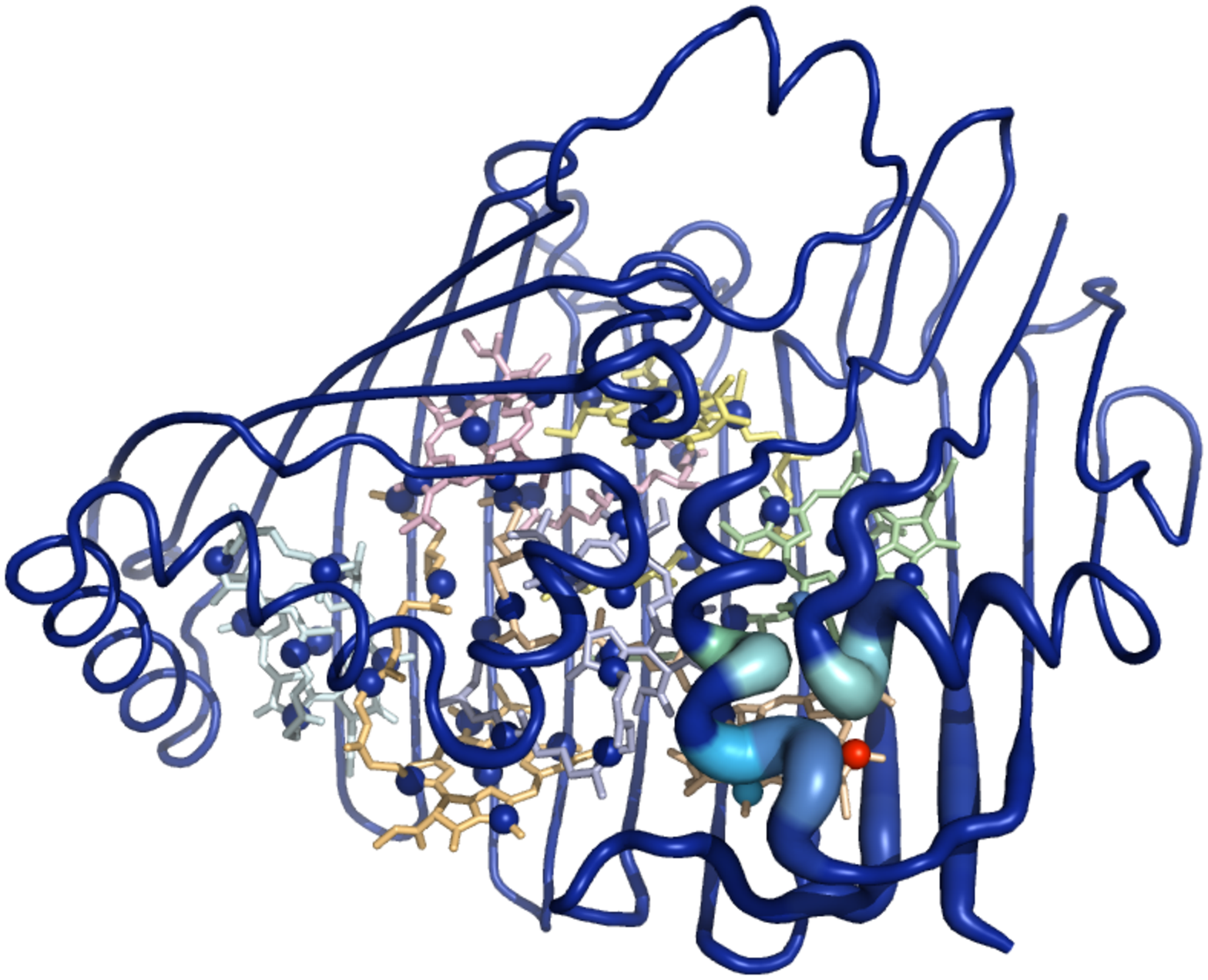}
\caption{Monomer 3 of the FMO complex coloured according to the average displacement during the MD simulation from 100 ps to 1 ns.}
\label{rmsd}
\end{figure}

In order to study the protein's dynamics we excite the highest frequency NM with a total energy of $40~\mathrm{kcal/mol}$ and then simulate the molecular dynamics, as described in the methods section. The highest frequency NM is chosen following work in the literature which suggests that localised DB modes can emerge from a subset of high frequency NMs \cite{Juanico2007,Luccioli2011}. In nature, these modes might be excited by thermal fluctuations or from the energy dissipation resulting from light absorption and recombination processes. We set $k_2=10~\mathrm{kcal/mol/}\text{\AA}^2$ and $k_4=10~\mathrm{kcal/mol/}\text{\AA}^4$, following Piazza et al. \cite{Juanico2007, Piazza2014}. Although the highest frequency NM is split equally across the three monomers, as described above, after 100 ps of the molecular dynamics simulation 72.8\% of the energy is found on monomer 3. In Fig. \ref{rmsd} we plot monomer 3 of the FMO complex coloured according to the average displacement during the MD simulation from 100 ps to 1 ns. We observe considerable localisation, in particular pigment 3 exhibits the largest displacement and we also observe substantial vibrational motion of helices 7 and 8.

\begin{figure}[ht]
\centering
\includegraphics[width=85mm]{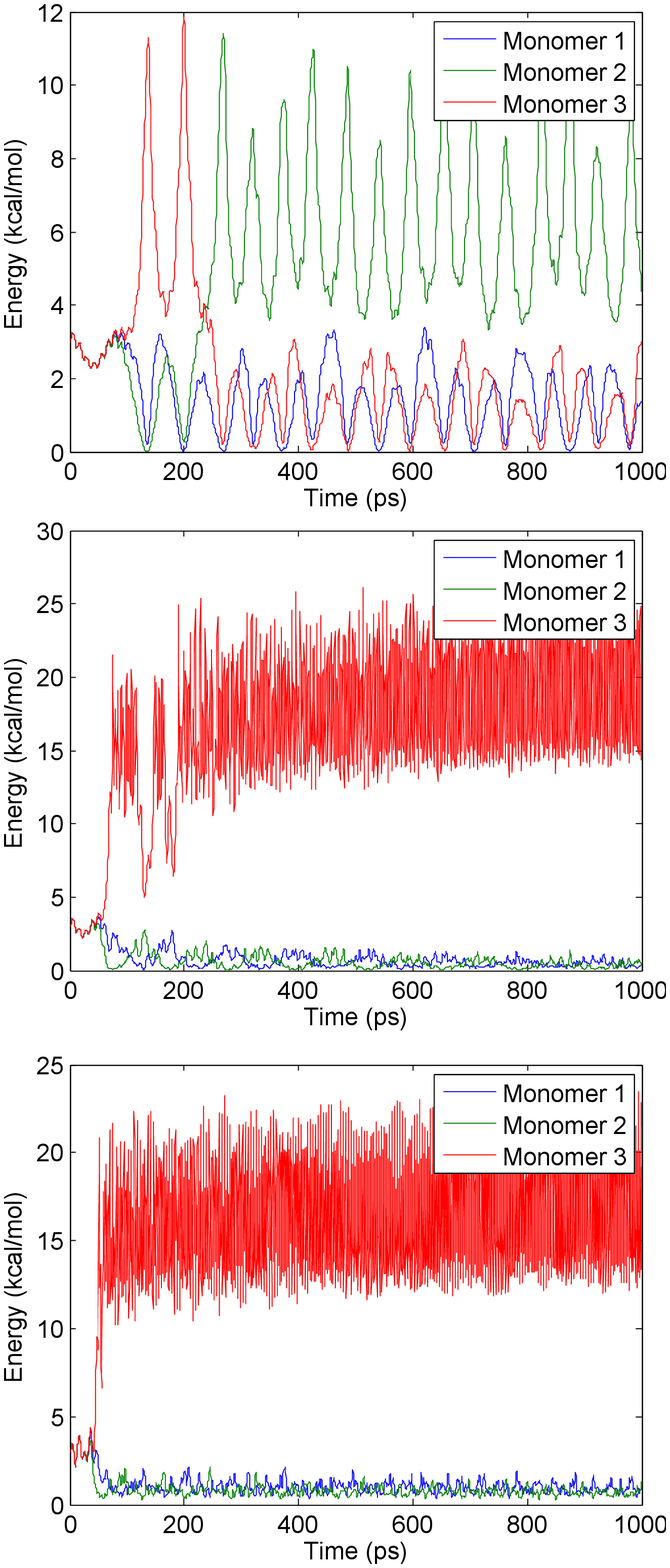}
\caption{The total energy on pigment 3 on each monomer over the first 1 ns following excitation of the highest frequency NM. Results are shown for $k_4=5$, 10 and 15$~\mathrm{kcal/mol/}\text{\AA}^4$ in \textbf{a}, \textbf{b} and \textbf{c}, respectively.}
\label{differentk}
\end{figure}

To study the localisation process in more detail, in Fig. \ref{differentk} we plot the total energy on pigment 3 on each monomer over time, for three different values of anharmonicity: $k_4=5, 10~\mathrm{and}~15~\mathrm{kcal/mol/}\text{\AA}^4$. As the anharmonicity increases, the localisation of energy on pigment 3 occurs more quickly, and vice versa. For $k_4=5~\mathrm{kcal/mol/}\text{\AA}^4$, we observe energy transfer between pigment 3 molecules on the three monomers (note that this continues at later times- for example after 1300 ps the energy is transferred to pigment 3 on monomer 1). Increasing/decreasing the excitation energy has a similar effect on the dynamics to increasing/decreasing the anharmonicity, for further details see Supplementary Fig. 2. Crucially, the localisation we observe around pigment 3 on a single monomer appears to be a fairly generic phenomenon, which occurs over a broad range of excitation energies and anharmonicity values. The energy remains localised for long times relative to excitation energy transfer in the FMO complex. Overall, the remarkable ability of the complex to harvest energy from multiple spatial locations and localise it on and around the core pigments could have important consequences for light harvesting processes. Pigments 3 and 4 have been shown to act as the energy sink for FMO and funnel energy excitations to the reaction centre \cite{Adolphs2006,Muh2007}, therefore vibrational energy transfer to this region is particularly interesting. Helices 7 and 8 have also been shown to be important in directing energy transport towards pigments 3 and 4 (the helix dipoles red shift the site energies) \cite{Muh2007}. We note that single localised modes have been shown to affect energy transfer \cite{Thorwart2015}, but we do not explore this further here.

We calculate the power spectrum of the x-displacement of the node in pigment molecule 3 with the highest energy (on the monomer with the highest energy), taken over a 300 ps period beginning after 300 ps of the molecular dynamics (not shown). For $k_4=10~\mathrm{kcal/mol/}\text{\AA}^4$, the main peak has a frequency of $104.8~\mathrm{cm}^{-1}$, which is approximately $5.4~\mathrm{cm}^{-1}$ higher than the frequency of the highest energy normal mode at $99.4~\mathrm{cm}^{-1}$ (similar results were obtained for displacements in the y and z directions). From work by Juanico and Piazza \cite{Juanico2007,Piazza2008}, this gap between the top of the harmonic spectrum and the frequency of the oscillations observed suggests the presence of a DB. The strong localisation of energy discussed above is also typical of DBs. For $k_4=5~\mathrm{kcal/mol/}\text{\AA}^4$ the main peak in the power spectrum shifts down to $100.1~\mathrm{cm}^{-1}$. When $k_4=15~\mathrm{kcal/mol/}\text{\AA}^4$, we find the main peak is at the higher frequency of $106.7~\mathrm{cm}^{-1}$ (in this case we take the trajectory between 100 ps and 400 ps).

In Fig. \ref{differentk} we observe both high and low frequency oscillations in the energy on pigment 3 over time. By taking the Fourier transform of the total energy on pigment 3 between 300 and 1300 ps, we find that the main high frequency peaks lie at 200.3, 210.9 and $215.4~\mathrm{cm}^{-1}$ for $k_4=5$, 10 and $15~\mathrm{kcal/mol/}\text{\AA}^4$ respectively. In each case this is approximately twice the frequency of the main peak in the power spectrum of the trajectory. We also observe low frequency oscillations, corresponding to peaks in the Fourier transform of the total energy on pigment 3 at 0.6, 6.3 and $7.1~\mathrm{cm}^{-1}$ for $k_4=5$, 10 and $15~\mathrm{kcal/mol/}\text{\AA}^4$ respectively. These low frequency oscillations are a result of the nonlinear term in equation \ref{eqn:PE}, however their exact origin remains unclear and requires further investigation. Picosecond timescales are known to be important for energy transfer in FMO, for example experimental observations by Zigmantas et al. \cite{Zigmantas2016} suggested a 17 ps timescale for excitation energy transfer between the lowest energy FMO state and the reaction centre. Low frequency oscillations such as those observed prominently in Fig. \ref{differentk} a could therefore provide alternative energy transfer pathways, perhaps enabling `trapped' energy to be redistributed.

Similar results are obtained when other high frequency NMs are excited, namely the energy becomes localised on a specific pigment or residue with high connectivity and a DB forms with a frequency which lies above the top of the harmonic spectrum. For example, when the fourth highest frequency NM is excited the energy becomes localised on residue GLY 231, which is found in the core of the complex near to helix 4 (the second and third highest frequency NMs are similar to the highest frequency NM, due to symmetry). Note that in this case we excite the complex with $E_0=50$ kcal/mol since $E_0=40$ kcal/mol does not lead to localisation on any single node within the first 1 ns. This is in contrast to simulations where low energy NMs are excited and the energy is dissipated across the protein over time. Examples of these dynamics are shown in the SI.

Having shown the intriguing possibility of nonlinear DB modes in FMO, we now turn to the effect which these modes might have on the optical spectra. In particular, we consider the linear absorption, linear dichroism and circular dichroism spectra, which are frequently used to examine molecular systems. As discussed in the methods section, we use the ab initio pigment energies and couplings obtained by Cole et al. \cite{Cole2012} to calculate the spectra, however crucially we also include a single discrete mode in our calculations in order to approximate the effect of a DB. This mode is treated at the same level as the electronic states. Above we showed that exciting the highest frequency NM leads to large displacements of pigment 3 as well as helices 7 and 8. We are not aware of any studies explicitly studying the variation in pigment site energies with helix displacements, but the proximity of helices 7 and 8 to pigments 4 and 7 suggests that their site energies are also likely to be modulated by the DB. Other pigments are not expected to be affected. Therefore we couple the mode to pigments 3, 4 and 7 only. In practice the strength of the coupling to different pigments may vary but as a first approximation we set the Huang Rhys factor S=0.3 for all three pigments. %since our analysis above shows that the site energies of those pigments are the most likely to be modulated by non-linear protein dynamics.

The frequency of the DB is unknown (it is determined by $k_2$ and $k_4$ in our model), but from the literature high frequency protein normal modes (excluding intramolecular pigment modes) might be expected to lie between approximately $100$ and $200~\mathrm{cm}^{-1}$ \cite{Juanico2007,Acbas2014,Piazza2008}. Given this uncertainty, in our calculations we consider two mode frequencies as examples: $\omega=100$ and $180~\mathrm{cm}^{-1}$. We note that a mode has been observed experimentally with $\omega=180~\mathrm{cm}^{-1}$ \cite{Wendling2000}. This mode is generally considered to be a pigment mode \cite{Wendling2000,Thorwart2015}, however as far as we are aware the intriguing possibility that this mode could be at least in part due to protein vibrations has not been ruled out. Interestingly, work by Leitner et al. suggested that protein modes might enhance chromophore vibrations in photoactive yellow protein \cite{Leitner2007}. Note that we require $k_2=30~ \mathrm{kcal/mol/}\text{\AA}^2$ and $k_4=30~\mathrm{kcal/mol/}\text{\AA}^4$ for the top of the harmonic spectrum to lie around $180~\mathrm{cm}^{-1}$ and previous work showing the high level of stiffness in FMO suggests that this is not unreasonable \cite{Fokas2014}. The dynamics following excitation of the highest frequency NM with $k_2=30~ \mathrm{kcal/mol/}\text{\AA}^2$ and $k_4=30~\mathrm{kcal/mol/}\text{\AA}^4$ are qualitatively similar to those obtained previously, leading to energy transfer to pigment 3, as shown in the SI (in this case we used $E_0=120$ kcal/mol to obtain localisation).

\begin{figure}[ht]
\centering
\includegraphics[width=180mm]{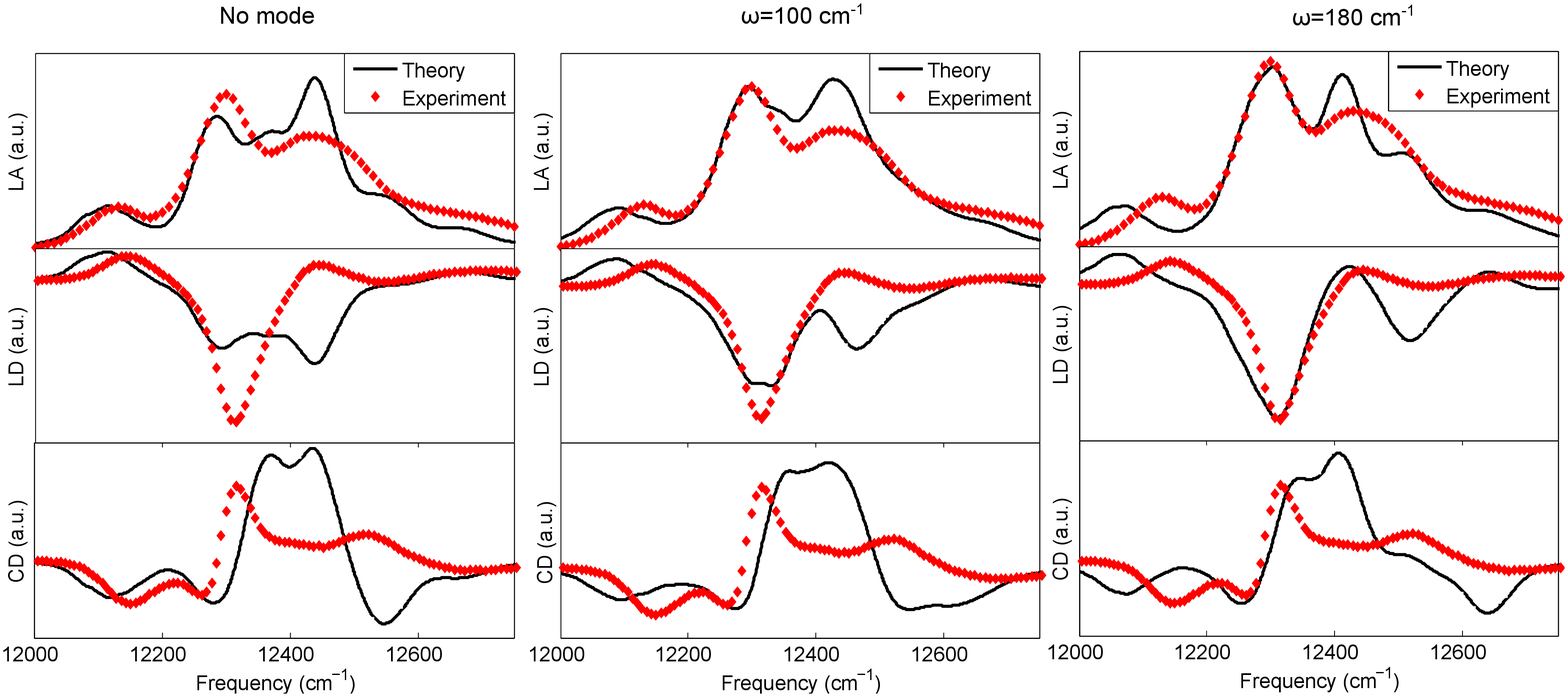}
\caption{LA, LD and CD spectra for calculations with no mode and for modes with $\omega=100$ and $180~\mathrm{cm}^{-1}$ (black lines). Experimental spectra are shown in red for comparison \cite{Engel2011}.}
\label{opticalspec}
\end{figure}

Figure \ref{opticalspec} plots optical spectra for $\omega=100~\mathrm{cm}^{-1}$ and $\omega=180~\mathrm{cm}^{-1}$ coupled to pigments 3, 4 and 7. We also plot the optical spectra with no mode (as calculated by Cole et al. \cite{Cole2012}), for comparison. Spectra with modes coupled to all pigments (not just pigments 3, 4 and 7) are shown in Supplementary Fig. 8. All calculated spectra are compared to experimental results for \textit{P. aestuarii} obtained at 77K \cite{Engel2011} (plotted in red). We note that whilst in equation \ref{eq:Hamiltonian} we have assumed linear coupling to the mode, when an anharmonic energy shift was included we did not notice any substantial spectral changes.

For both $\omega=100~\mathrm{cm}^{-1}$ and $\omega=180~\mathrm{cm}^{-1}$, linear absorption and linear dichroism spectra obtained by coupling a mode to pigments 3, 4 and 7 are closer to the experimental results than results obtained with no mode or a mode with equal coupling to all pigments. In particular, the peak around $12300~\mathrm{cm}^{-1}$ in the linear absorption, which was underestimated in calculations with no mode becomes more prominent. Meanwhile the peak around $12450~\mathrm{cm}^{-1}$, which was overestimated in the calculations with no mode, is reduced. In the linear dichroism spectrum, the peak at $12300~\mathrm{cm}^{-1}$ was  underestimated in the original calculations. When the mode is included this peak is much deeper. The linear dichroism spectra around $12400~\mathrm{cm}^{-1}$ are also much closer to experimental results with the mode, particularly when $\omega=180~\mathrm{cm}^{-1}$. The inability of mode coupling to all pigments to capture these effects, as shown in the SI, highlights the potential importance of localised non-linear protein oscillations. Note that the coupling of the mode to pigment 3 lowers the energy of the lowest energy exciton, which red shifts the low energy parts of the spectra, slightly worsening the agreement with the experimental results in that region. This discrepancy could be caused by an error in the ab initio pigment 3 site energy, which is lower than the result obtained by Adolphs et al. for example \cite{Adolphs2006}. Various approximations were employed in the calculation of ab initio site energies \cite{Cole2012}. Alternatively, the mode might couple more strongly to pigments 4 and 7 than to pigment 3, perhaps because the DB mode which emerges from exciting the highest frequency NM is not representative of the average DB coupling across all complexes or due to other effects not captured by the relatively simple nonlinear network model. Further work is required to understand the relative importance of these sources of error and also to elucidate the circular dichroism results, which are not substantially improved by incorporating the mode.

%Up to three levels of \textbf{subheading} are permitted. Subheadings should not be numbered.
%\subsection*{Subsection}
%\subsubsection*{Third-level section}
%Topical subheadings are allowed.

%\section*{Discussion}
%The Discussion should be succinct and must not contain subheadings.

\section*{Conclusions}

To our knowledge, this work is the first to apply the nonlinear network model to study the archetypal light-harvesting complex, FMO. This computationally inexpensive approach allows us to investigate FMO's collective protein modes, which have been implicated in the energy transfer processes, at a microscopic level. Our model also incorporates an anharmonic term which enables us take a first step towards capturing the nonlinear effects that previous work shows can play a role in protein dynamics \cite{Levy1982,Hayward1995,Austin2000}.

We find that the high frequency NMs can be delocalised across the trimer and exhibit strong components on the core pigments and on certain residues with high connectivity. When high frequency NMs are excited, we observe energy localisation in biologically relevant areas of the protein. For example exciting the highest frequency NM leads to energy localisation around pigment 3 and helices 7 and 8, which are expected to modulate the site energies of pigments 4 and 7. By studying the power spectrum we show that a DB mode has been formed and similar results can be obtained following excitation of other high frequency NMs. These non-linear DB modes offer the potential for intriguing alternative energy transfer routes. In particular they could play a role in localising and storing excess energy around pigments 3 and 4 which are known to funnel energy excitations towards the reaction centre. Future work should focus on the details of these processes, both experimentally and by establishing more accurate theoretical models. The ability of DBs to transfer energy around the protein could also lead to a non-equilibrium environment which is temporally spatially dependent; in other words the environment has a spatial dependence which changes over time. These dynamics might offer alternative ways to co-ordinate excitation energy transfer. This type of environment has received little attention to date and could open up exciting possibilities for device design.

Having established the existence of potentially biologically important DBs in our system, we investigated the effect which these localised modes might have on the optical spectra which were calculated by Cole et al. \cite{Cole2012} using ab initio site energies and couplings. To do this we approximated the DB as a single mode, localised on pigments 3, 4 and 7 with example frequencies of $\omega=100~\mathrm{or}~180~\mathrm{cm}^{-1}$. We incorporated this mode into optical spectra calculations on an equal footing with the electronic energy levels. Substantial improvements in the LA and LD spectra were obtained, particularly around the $12300-12450~\mathrm{cm}^{-1}$ region. We note that this is in contrast to the effect of incorporating very low frequency modes, which previous work found did not affect the spectra \cite{Fokas2014}. Nonetheless there is room for improvement, in particular the lowest energy exciton is lower than observed experimentally and the circular dichroism spectra still show  large differences from the experimental results. This work highlights the importance of ab initio calculations of the pigment energies and couplings and further work is needed to obtain more accurate ab initio site energies \cite{Zuehlsdorff2015}.

\bibliography{references}

\section*{Acknowledgements}

%Acknowledgements should be brief, and should not include thanks to anonymous referees and editors, or effusive comments. Grant or contribution numbers may be acknowledged.

We are grateful to Dugan Hayes (Engel Group, U. Chicago) for providing the experimental absorption spectrum data used for comparison in this work and Rienk van Grondelle and Markus Wendling for the experimental LD and CD spectra. We also thank Alexander Fokas for helpful discussions. A.W.C. and S.E.M. acknowledge support from the Winton Programme for the Physics of Sustainability. S.E.M. is also supported by an EPSRC doctoral training award. D.J.C. is supported by a Marie Curie International Outgoing Fellowship within the seventh European Community Framework Programme.

\section*{Author contributions statement}

S.E.M. ran the simulations, S.E.M., D.J.C. and A.W.C. analysed the results and all authors reviewed the manuscript.

\section*{Additional information}

The author(s) declare no competing financial interests.

\end{document}